\documentstyle[psfig]{l-aa}

\begin{document}

\thesaurus{05                
           (08.08.1;         
            08.05.2          
            08.12.3          
            11.19.4.NGC\,2006  
            11.19.4 SL\,538    
            11.13.1)         
            }                

\title{The cluster pair SL 538/ NGC 2006 (SL 537)
\thanks{Based on observations taken at the European Southern Observatory, La
  Silla, Chile, during time allocated by the MPIA, Heidelberg.}} 

\author{Andrea Dieball\inst{1}
        \and
        Eva K. Grebel\inst{2,3}
       } 

\offprints{A. Dieball, adieball@astro.uni-bonn.de}
\institute{Sternwarte der Universit\"{a}t Bonn, Auf dem H\"{u}gel 71, 
           D--53121 Bonn, F.R. Germany, 
           adieball@astro.uni-bonn.de \and
           Astronomisches Institut der Universit\"{a}t W\"{u}rzburg, Am Hubland,
           D--97074 W\"{u}rzburg, F.R. Germany \and
           UCO / Lick Observatory, University of California, Santa Cruz, CA
           95064, USA}
  
\date{Received date, accepted date}

\maketitle

\markboth{Dieball \& Grebel: The cluster pair SL\,538 \& NGC\,2006}{}

\begin{abstract}

We studied in detail the binary cluster candidates SL\,538 and NGC\,2006 in
the Large Magellanic Cloud (LMC).  
This cluster pair is located in the northwestern part of the large OB
association LH\,77 in supergiant shell LMC\,4. A third star
cluster, KMHK\,1019, is located within $5\arcmin$ from the cluster pair. 
Our study is the first age determination of SL\,538 and NGC\,2006 that is
based on CMDs. We derived an age of $18\pm2$ Myr for SL\,538, $22.5\pm2.5$ Myr
for NGC\,2006, and $16$ Myr for KMHK\,1019. Thus the three clusters 
are (nearly) coeval. We identified Be star candidates and find the same
ratio N(Be)/N(B) for the components of the binary cluster (12\%) while the
amount of Be stars detected in KMHK\,1019 (5\%) and in the surrounding field
(2\%) is considerably lower. Since Be stars are usually rapid rotators this
may indicate intrinsically higher rotational velocities in the
components of the cluster pair. Also the IMF derived from the CMDs shows the
same slope for both SL\,538 and NGC\,2006 and is consistent with a Salpeter
aIMF. An estimation of the cluster masses based on the IMF slopes showed that
both clusters have similar total masses. These findings support joint,
near-simultaneous formation of the cluster pair in the same giant molecular
cloud.  
 
\keywords{Hertzsprung-Russel (HR) diagram -- stars: emission line, Be --
          stars: luminosity function, mass function -- star clusters: 
          individual: NGC\,2006 and SL\,538 --
          Magellanic Clouds}

\end{abstract}


\section{Introduction}
 
The existence of gravitationally bound pairs of star clusters is important for
the understanding of formation and evolution of star clusters. Since the 
probability of tidal capture of one cluster by another one is very small
(Bhatia et al. 1991), we
can assume that the components of a true binary star cluster have a common 
origin. Star clusters form in giant molecular clouds (GMCs) (e.g. Elmegreen \&
Elmegreen 1983), but the details of cluster formation are not yet well
understood. If the components of a cluster pair or multiple cluster formed 
simultaneously or sequentially 
in the same fragmenting GMC they should have similar properties like age,
metallicity and stellar content.      

In the Milky Way only a few binary clusters are known, though Lyng{\aa} \&
Wramdemark (1984) suggest the common origin of a group of six Galactic open
clusters. Later Pavloskaya \& Filippova (1989), and more recently Subramaniam
et al. (1995), propose the existence of more possible Galactic binary clusters
and cluster complexes.  

The apparent lack of binary clusters in our own Galaxy may be explained in
different ways. Subramaniam et al. (1995) argue that since we are looking at
the Galaxy from inside, double clusters may be harder to detect than in the
distant Magellanic Clouds, where binary clusters can easily be detected due to
the closeness of their projected positions on the sky. The distance to the 
Galactic clusters must also be taken into account, but only approximately 
400 of 1400 open clusters have known distances (Lyng\aa\ 1987). Subramaniam et
al. (1995) found 16 Galactic binary cluster candidates on the base of the
Lyng\aa\, catalogue, which corresponds to 8 \% of the investigated number of 
clusters. {}From this they conclude that binary clusters in the Milky Way may
not be uncommon. 

The evolution of a gravitationally bound pair of star clusters depends on
the interaction between the components as well as on the tidal forces of the
parent galaxy. If the tidal field is strong, the binary system will not
survive for long but soon will get disrupted. {}From some preliminary
considerations Innanen et al. (1972) conclude that due to stronger tidal
forces in the Milky Way a binary cluster will execute only a fraction of a 
single orbit around the barycentre before its components are detached, 
but it will survive for several orbits in the less dense, less massive
Magellanic Clouds.  
Surdin (1991) came to the same conclusion, especially for massive clusters. 
The investigation of binary clusters may help to evaluate the tidal field 
of the parent galaxy. 

{}Fujimoto \& Kumai (1997) suggest that globular and populous star clusters
form through strong collisions between massive gas clouds in 
high-velocity random motion. Shear and momentum of oblique cloud-cloud
collisions lead to break-up into compressed sub-clouds revolving around each
other, which may form binary or multiple clusters.  Binary star clusters are
expected to form more easily in galaxies like the Magellanic Clouds with 
high-velocity random gas motions, whereas in the Milky Way such large-scale
high-velocity random motions are lacking.

Bhatia \& Hatzidimitriou (1988), Hatzidimitriou \& Bhatia (1990), and Bhatia et
al. (1991), have surveyed the Magellanic Clouds in order to catalogue the 
binary cluster candidates. The maximum projected centre-to-centre separation 
of the components of a pair was chosen to be 18 pc, which corresponds to 
$\simeq 1\farcm3$ in the LMC. A binary cluster with larger separation may
become detached by the external tidal forces while shorter separations may
lead to mergers (Sugimoto \& Makino 1989, and Bhatia 1990).    
In these studies 69 pairs in the LMC and 9 pairs in the SMC were identified. 
Two clusters may appear to be a binary cluster due to chance line-up while 
in fact being at different distances within the Magellanic Clouds and not
gravitationally bound to each other. The number of chance-pairs of objects
uniformly distributed in space can be estimated with a formula presented by
Page (1975). Taking into account also a non-uniform distribution of star
clusters (at least for the LMC), Bhatia \& Hatzidimitriou (1988) and
Hatzidimitriou \& Bhatia (1990) found that statistically 31 pairs in the LMC
and 3 pairs in the SMC could be explained due to mere chance line-up. As
considerably more pairs have been found, this strongly suggests that at least
a certain amount of them must be true binary clusters.
 
While it is difficult to measure true distances between apparent binary
clusters an analysis of their age and stellar content can give clues to a
possible common origin.

The star cluster pair NGC\,2006 (also known as SL\,537) and SL\,538 is located
in the northwestern part of the OB association LH\,77 in supergiant shell
LMC\,4, and has a projected centre-to-centre separation of $54\farcs67$
corresponding to 13.3 pc. This double cluster has already been the subject of
investigations concerning its binarity. Bhatia (1992) and Bica et al. (1996)
found from integrated photometry that both clusters have the same
age. Kontizas et al. (1993) analyzed the stellar content and the cores of the
components using low resolution objective prism spectra and integrated IUE
spectra. They suggested this cluster pair constitutes a 
true binary cluster, which moreover may merge in some $10^{7}$ years. This
raises another question, which has also been discussed in Bhatia \&
MacGillivray (1988): could mergers of former binary star clusters be
responsible for at least some of the blue populous clusters in the LMC?

We investigated the double cluster NGC\,2006 and SL\,538 in an attempt to find
further affirmation -- or disaffirmation -- of the binarity of the two
clusters. We analyze the star density in the clusters and the surrounding
field (Section 3). 
For the first time we derive ages for these clusters from isochrone fits to 
colour-magnitude diagrams (CMDs) (Section 4), which is a much more reliable
age determination than using integrated photometry. In Section 5 we
investigate the content of Be stars in the clusters as well as in the
surrounding field. 
In Section 6 we give a summary and conclusions. 

\section{The data and the data reduction}

\begin{figure}
\centerline{
\psfig{figure=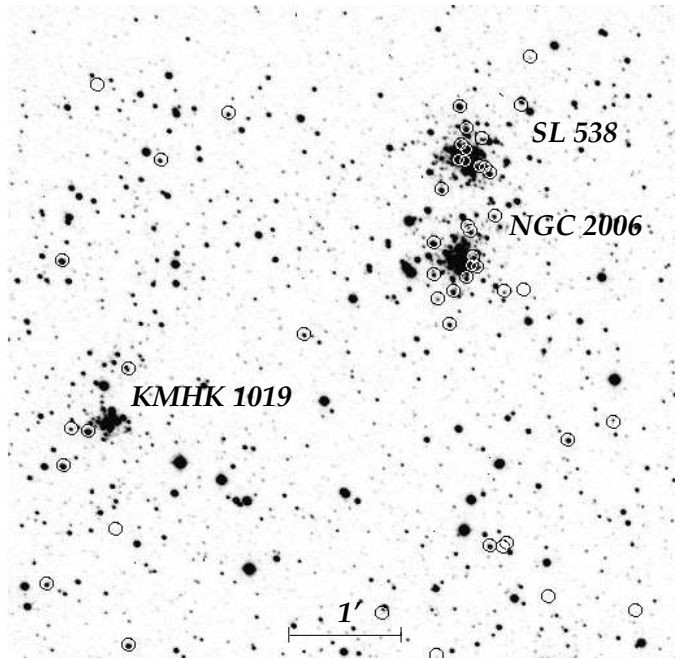,width=8.9cm,height=8.7cm}
}
\caption[]{$V$-image of SL\,538 and NGC\,2006,
  north is up and east to the left. The field of view of our image is $5\farcm8
  \times 5\farcm8$. To the south-east of the cluster pair a third star
  cluster is located: KMHK\,1019. The location of the Be star candidates (see
  Sect. 5) is marked with circles}
\label{sl538}
\end{figure}

An image of SL\,538 and NGC\,2006
is shown in Fig. \ref{sl538}. To the south-east of the cluster pair a third
star cluster is located: KMHK\,1019.  

The data have been obtained on 15 December 1992 with EFOSC 2 at the ESO/MPI
2.2 m telescope at La Silla. A $1024 \times 1024$ coated
Thomson THX31156 chip (ESO \#19) was used with a pixel scale of $0\farcs 34$
resulting in a field of view of $5\farcm8 \times 5\farcm8$.  All data were 
obtained with the standard Bessell $B$, $V$, $R$, Gunn $i$ and $H\alpha$ 
filters used at the 2.2 m telescope (see Table \ref{obs} for an observing log).

After standard image reduction with MIDAS, profile fitting photometry
was carried out with DAOPHOT~II (Stetson 1991) running under MIDAS.

The photometry was transformed using the Landolt standard fields
around PG\,0218+029, Rubin\,149, and SA\,98 (Landolt 1992) observed in the
same night. 

We applied the following transformation relations:

\begin{eqnarray*}
&&b-B = z_{B} + a_{B} \cdot X + c_{B} \cdot (B-V)\\
&&v-V = z_{V} + a_{V} \cdot X + c_{V} \cdot (B-V)\\
&&r-R = z_{R} + a_{R} \cdot X + c_{R} \cdot (V-R)\\
&&i-I = z_{I} + a_{I} \cdot X + c_{I} \cdot (V-I),
\end{eqnarray*} 

where $X$ is the mean airmass during observation, capital letters
represent standard magnitudes and colours, and lower-case 
letters denote instrumental magnitudes after normalizing to
an exposure time of 1 sec. The resulting colour terms $c_{i}$, zero
points $z_{i}$ and atmospheric extinction coefficients $a_{i}$ are:

\begin{eqnarray*}
&& z_{B} = 2.593\pm0.049 \quad\mbox{mag} \\
&& a_{B} = 0.140\pm0.039 \quad\mbox{mag}\\
&& c_{B} = -0.257\pm0.007 \\
&& z_{V} = 1.271\pm0.038 \quad\mbox{mag}\\
&& a_{V} = 0.135\pm0.030 \quad\mbox{mag}\\
&& c_{V} = -0.063\pm0.006 \\
&& z_{R} = 1.170\pm0.050 \quad\mbox{mag} \\
&& a_{R} = 0.081\pm0.039 \quad\mbox{mag}\\
&& c_{R} = 0.007\pm0.015 \\
&& z_{I} = 2.504\pm0.040 \quad\mbox{mag} \\
&& a_{I} = 0.073\pm0.032 \quad\mbox{mag}\\
&& c_{I} = 0.047\pm0.005 
\end{eqnarray*}

\begin{table}
\caption[]{\label{obs}Observing log}
\begin{tabular}{lcrc}
\hline
Object          & Filter&\multicolumn{1}{c}{Exp.time}& Seeing  \\
                &       & [sec]                      &[\arcsec]\\
\hline
SL\,538/NGC\,2006 &$B$      & 360                        & 1.5 \\
SL\,538/NGC\,2006 &$B$      & 90                         & 1.5 \\
SL\,538/NGC\,2006 &$V$      & 150                        & 1.3 \\
SL\,538/NGC\,2006 &$V$      & 20                         & 1.1 \\
SL\,538/NGC\,2006 &$R$      & 150                        & 1.3 \\
SL\,538/NGC\,2006 &$R$      & 30                         & 1.4 \\
SL\,538/NGC\,2006 &Gunn $i$ & 180                        & 1.3 \\
SL\,538/NGC\,2006 &Gunn $i$ & 20                         & 1.3 \\
SL\,538/NGC\,2006 &$H\alpha$& 900                      & 1.2 \\
\hline
\end{tabular}
\end{table}

\section{Stellar density in and around the clusters}

\begin{figure*}
\centerline{\hbox{
\psfig{figure=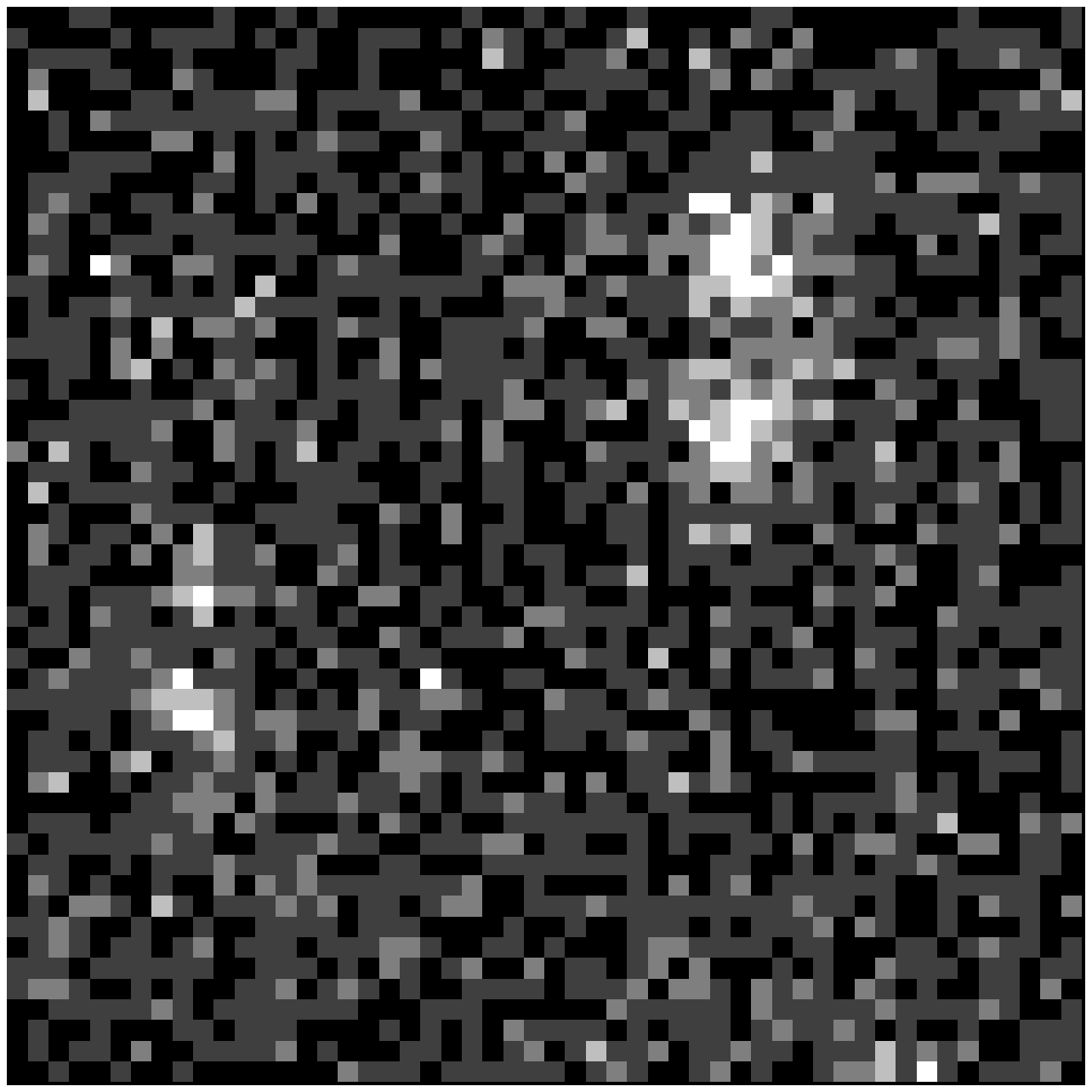,width=8.9cm,height=8.7cm}
\psfig{figure=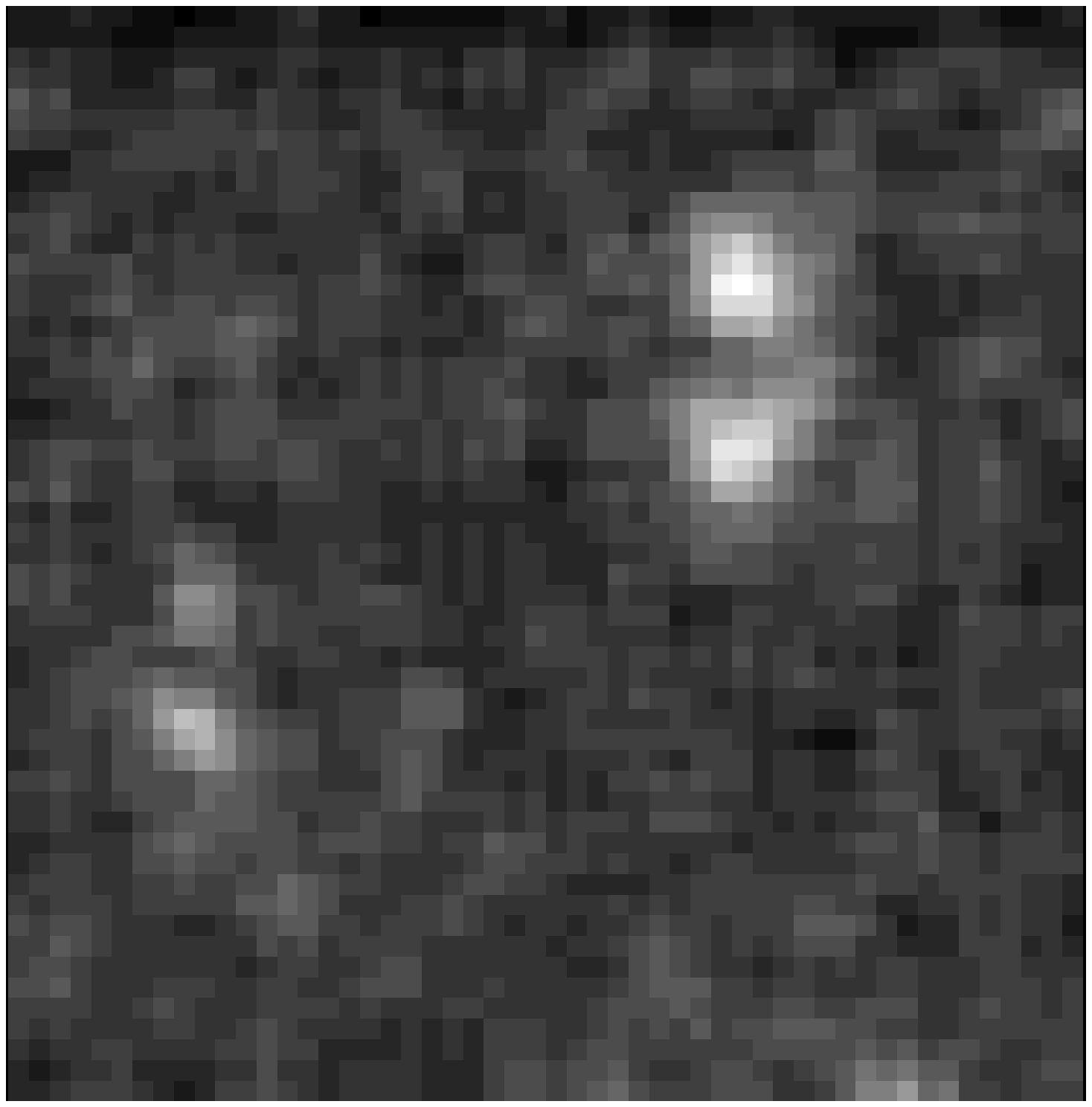,width=8.9cm,height=8.7cm}
}}
\caption[]{Left: Investigating the stellar surface density distribution we see
  enhanced stellar density between the two components while no enhancement
  between the binary cluster candidate and KMHK\,1019 can be seen. We plotted
  8 grey values, reaching from black (no star per cell) to white (7 stars per
  cell). Right: To make density structures and thus the stellar bridge between
  the two clusters better visible we applied a $3\times3$ average filter for
  image smoothing. Note that one cannot use this image for star counting as the
  pixel values are averaged to repress noise}
\label{sl538dens}
\end{figure*}

In order to investigate the stellar surface density distribution in and around
the clusters we subdivided the entire area into 52 square cells with 20 pixels
(corresponding to $6\farcs8$) length each.
We counted the number of stars found within each cell. In Fig. \ref{sl538dens}
(left) one cell corresponds to one pixel of the star density plot. 
The counting was carried out on the
ALLSTAR-output tables, which contain the coordinates of each measured star.
We see an enhanced star density between the cluster pair.
The apparent stellar bridge between the components contains as many as 25\%
of the stars counted in each of the clusters. To make density structures and
thus the stellar bridge between the two clusters better visible we applied
a $3\times3$ average filter
for image smoothing (see Fig.\ref{sl538dens}, right). Since we expect
especially high values in the stellar bridge compared to the surrounding
field, a median filter -- which does not 
consider very high or very low pixel values -- seems not to be appropriate for
our purpose. However, we also tried a median filter and found no differences
concerning the features of the resulting images.

No signs of an increased star density reaching from KMHK\,1019 
towards the binary cluster candidate can be seen. 

\begin{figure*}
\centerline{\hbox{
\psfig{figure=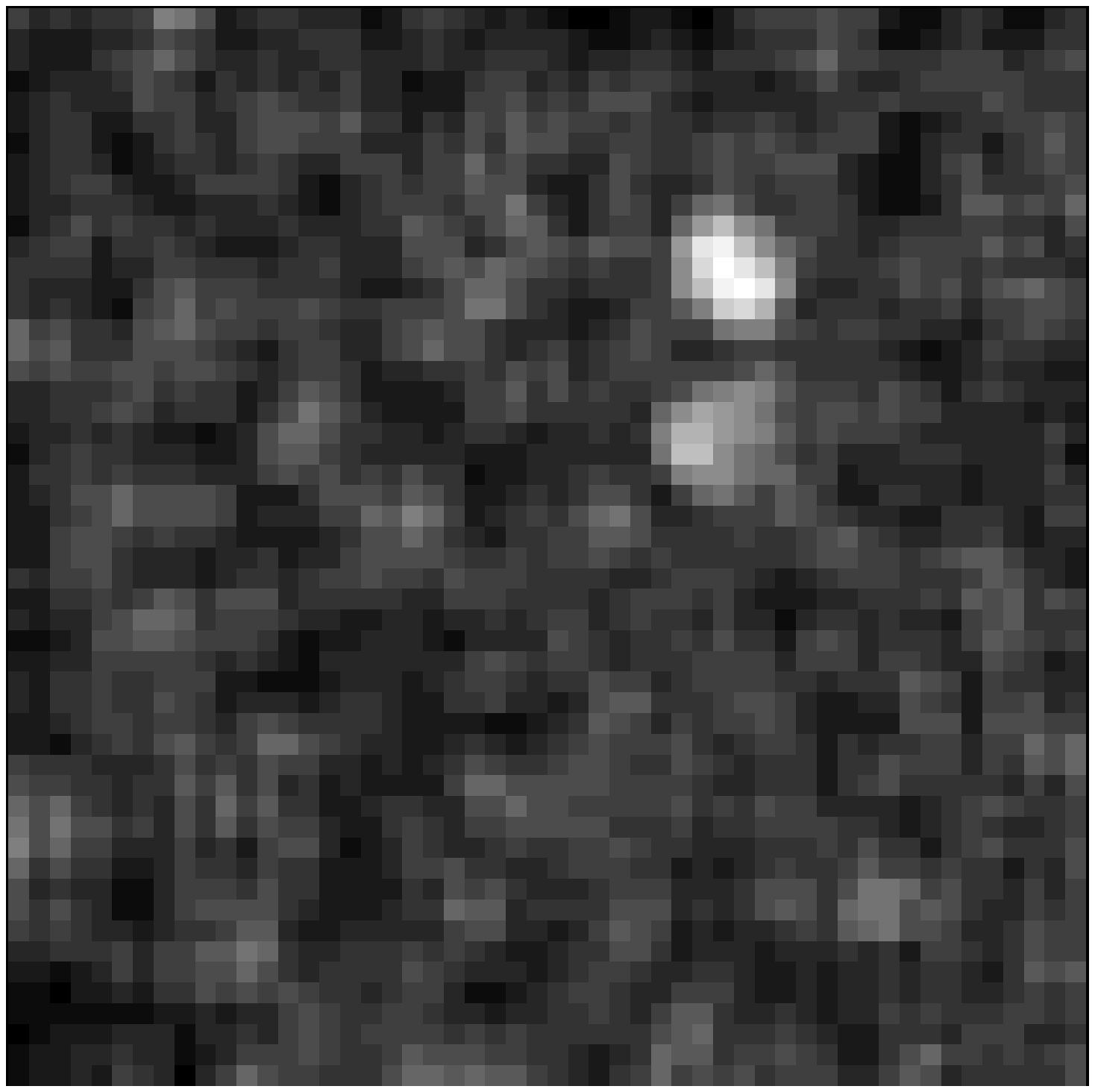,width=6cm,height=6cm}
\psfig{figure=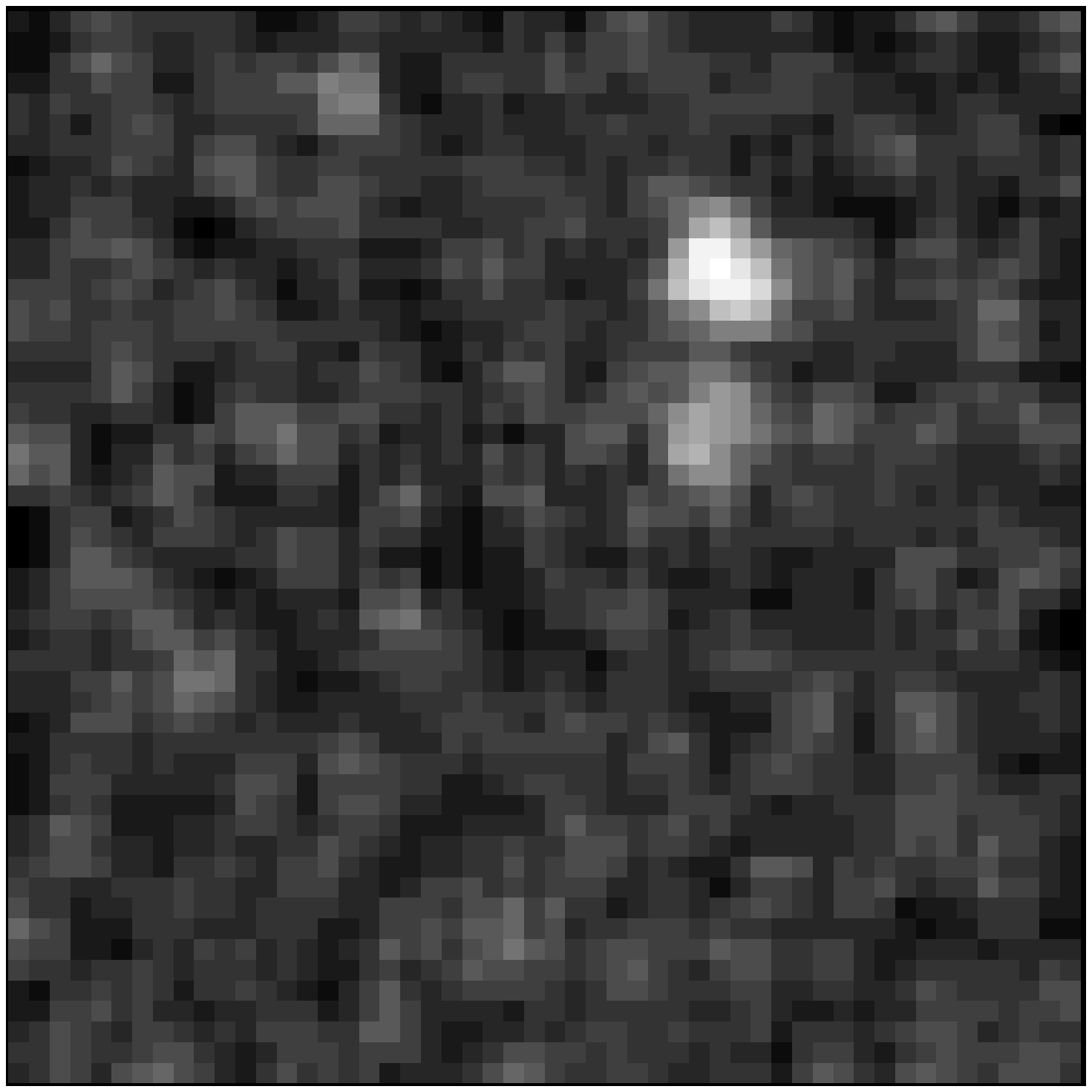,width=6cm,height=6cm}
\psfig{figure=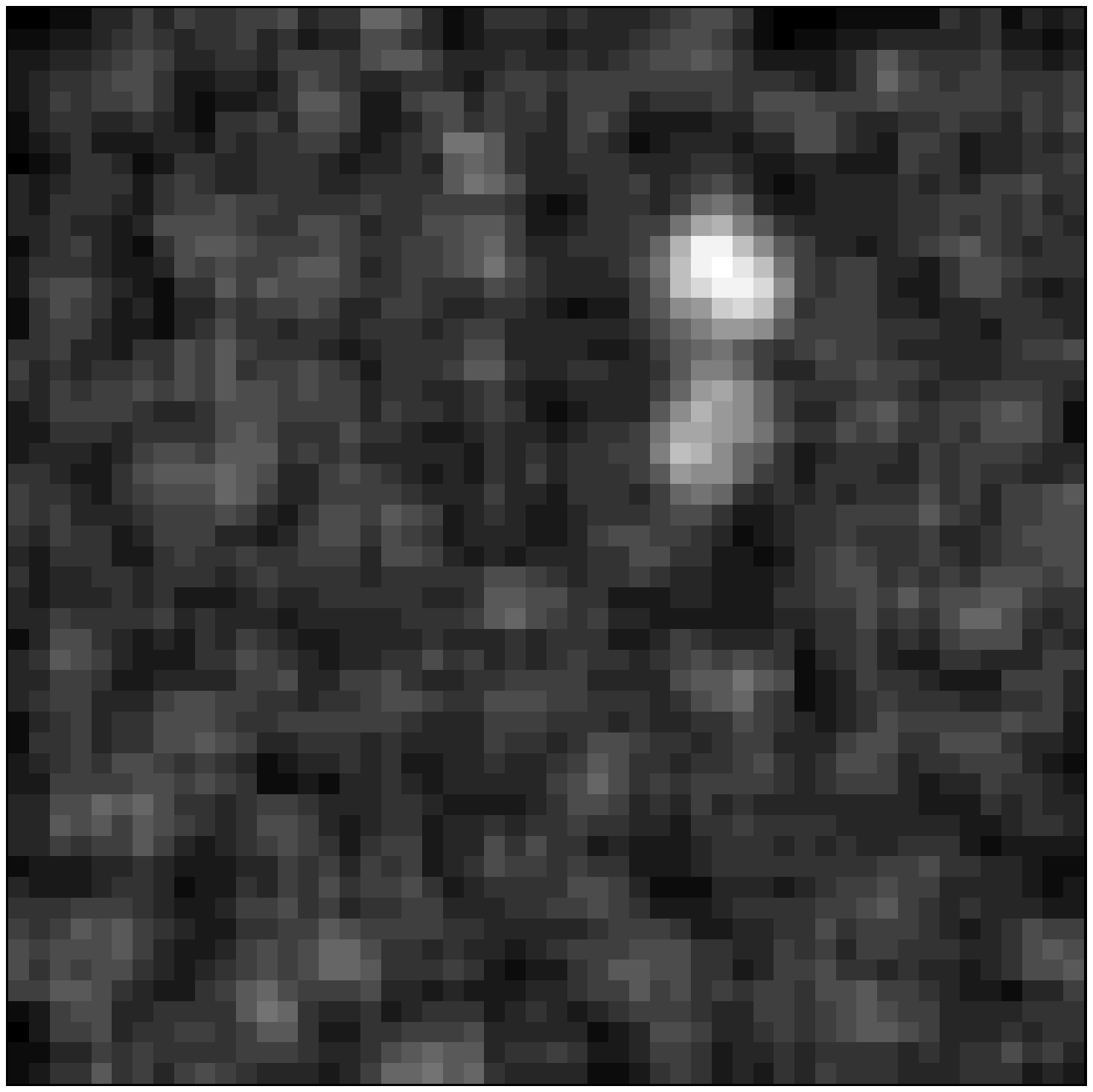,width=6cm,height=6cm}
}}
\caption[]{Star density plots based on artificial star experiments. Left: no
  bridge occurred between the star clusters, middle: a small bridge connects
  the components of the cluster pair, right: the stellar bridge is more
  pronounced. This last case occurred only three times out of 50
  star experiments and shows that there is a low likelihood of a
  pronounced stellar bridge produced by random fluctuations (6\%)}
\label{sldens}
\end{figure*}

We investigated the statistical probability of seeing such a stellar
bridge due to chance density fluctuations. 
For this purpose we removed all stars outside a 60 pixels radius
around the components of the cluster pair and carried out 50 artificial star
experiments with ADDSTAR running under DAOPHOT II. Each time we added the same
number of stars at random coordinates as had been removed initially.
Star density plots were created in the same way as described above. 
We found 12 plots out of 50 showing larger star density somewhere between --
and connecting -- the star clusters. In most
cases the stellar bridge connecting the components is very thin.
Only three plots showed
stellar bridges wider than one cell. In Fig. \ref{sldens} three examples of our
artificial star density plots are shown: no bridge, small bridge, and wide
bridge. Our artificial star experiments show that in one out of four cases a
stellar bridge may occur due to statistical density fluctuations in the field.
Thus, the probability of a stellar bridge which is not due to
statistical star density fluctuations does not reach a confidence level 
of 95 \%.   

\section{Deriving ages for the star clusters}

We derived ages of the clusters SL\,538, NGC\,2006, KMHK\,1019, and the
surrounding field star populations by comparing our CMDs with isochrones. 
The isochrones we used are based on the stellar models 
of the Geneva group (Schaerer et al. 1993).  

\subsection{Age determination}

\begin{figure*}
\centerline{
\psfig{figure=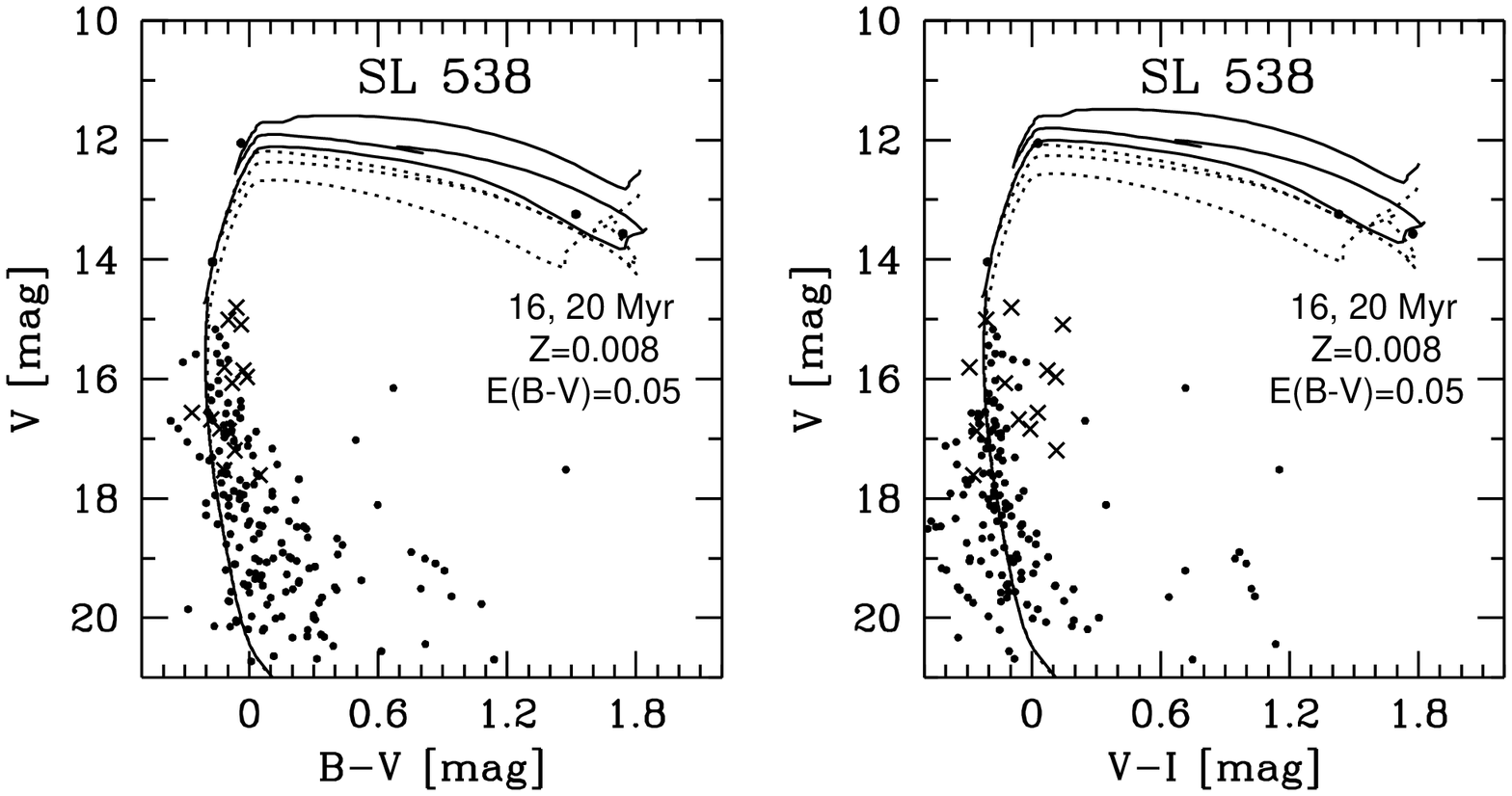,width=15cm,height=8cm,bbllx=15mm,bblly=96mm,bburx=202mm,bbury=210mm}}
\vspace{-0.5cm}
\caption[]{Colour-magnitude diagram of the star cluster SL\,538. We adopt an
  age of $18\pm2$ Myr. The derived age is supported by the best-fitting
  isochrones in all diagrams. Be star candidates (Sect. 5) are
  marked with crosses}
\label{cmdsl538}
\centerline{
\psfig{figure=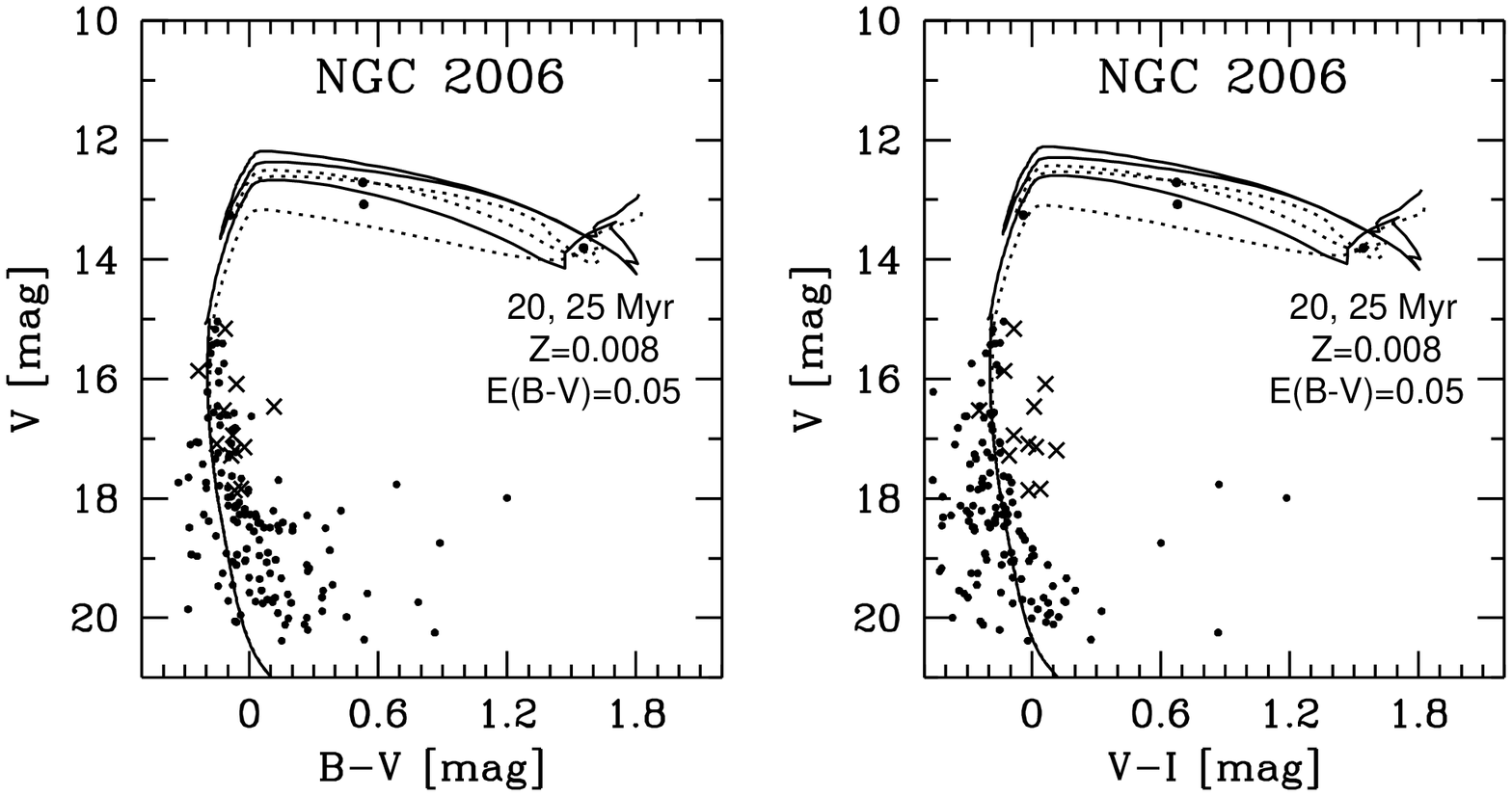,width=15cm,height=8cm,bbllx=15mm,bblly=96mm,bburx=202mm,bbury=210mm}}
\vspace{-0.5cm}
\caption[]{Same as above, but for the cluster NGC\,2006. The adopted age is
  $22.5\pm2.5$ Myr, which is confirmed by the isochrone fits in all
  colours}
\label{cmdn2006}
\end{figure*}

\begin{figure*}
\centerline{
\psfig{figure=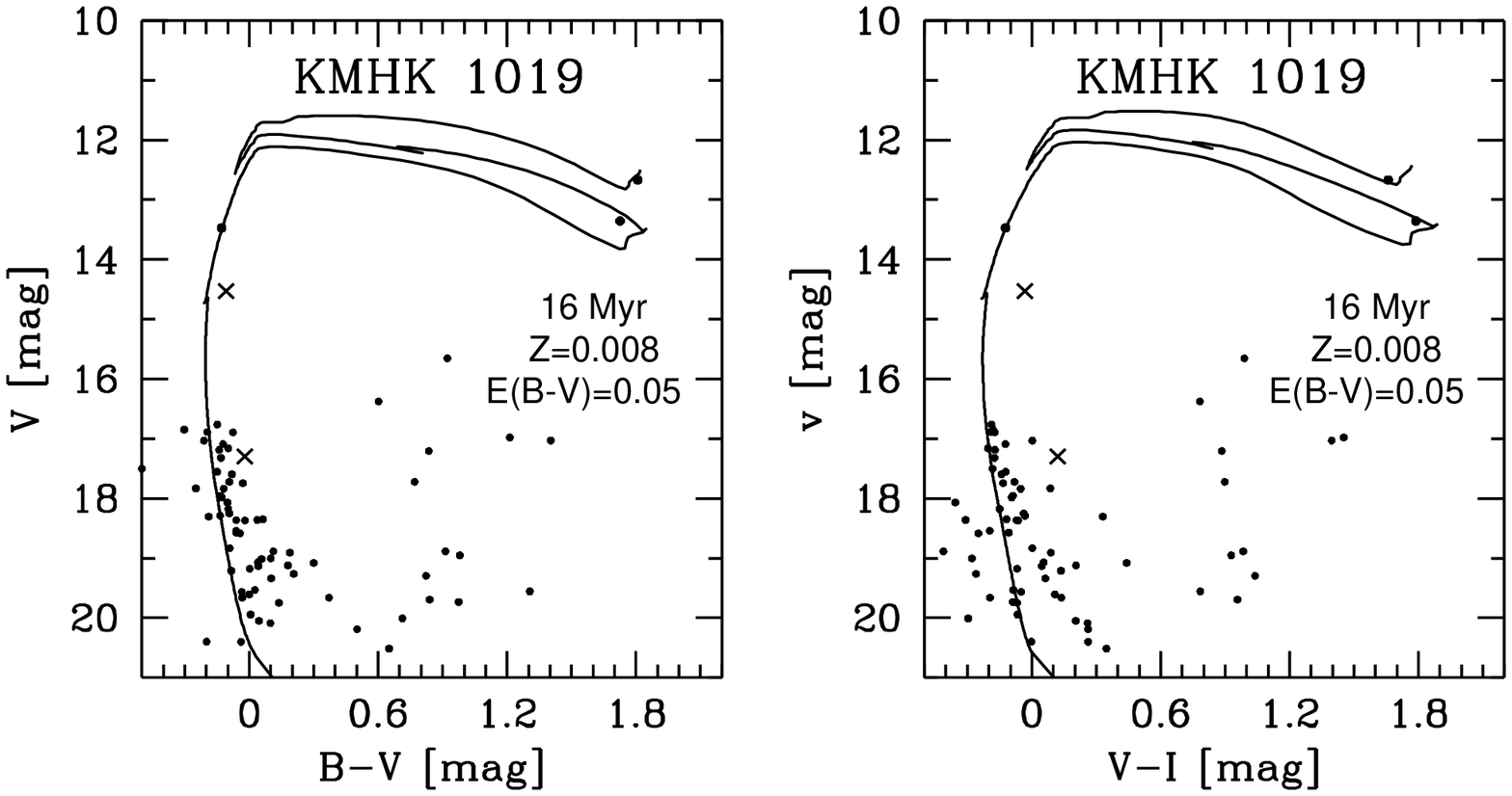,width=15cm,height=8cm,bbllx=15mm,bblly=96mm,bburx=202mm,bbury=210mm}}
\vspace{-0.5cm}
\caption[]{CMD of KMHK\,1019, the smallest of the three star clusters and with
  the lowest number of stars. The resulting age for the
  cluster from the best fitting isochrone is 16 Myr}
\label{cmdkmhk}
\centerline{
\psfig{figure=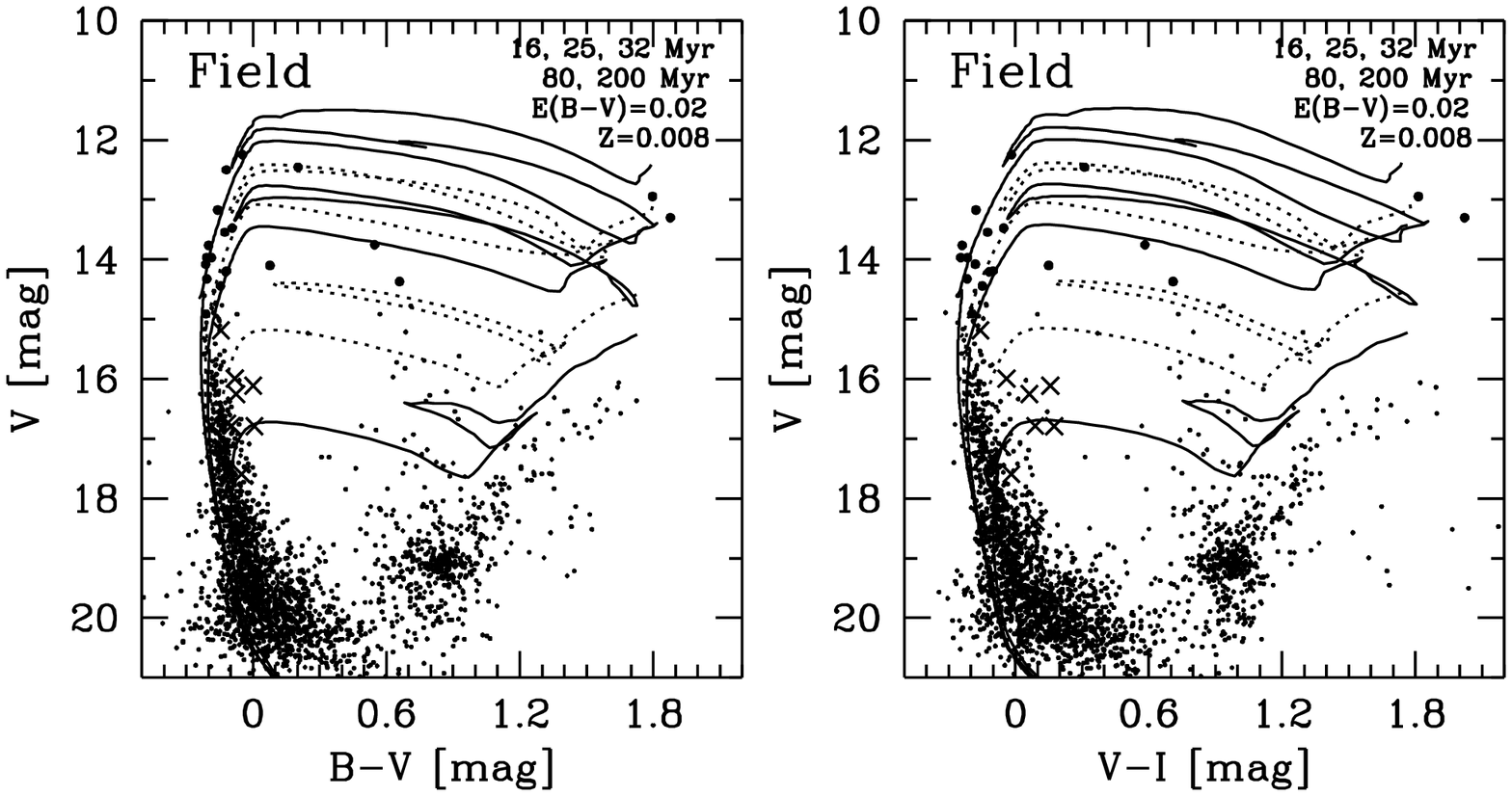,width=15cm,height=8cm,bbllx=15mm,bblly=96mm,bburx=202mm,bbury=210mm}}
\vspace{-0.5cm}
\caption[]{CMD of the surrounding field. This diagram comprises the mixture of
  ages of the various field populations: the blue main sequence and the
  supergiants represent the younger populations, the intermediate age
  populations show up through the RGB and the pronounced clump. Note the
  widening of the main sequence due to Be stars (see Sect. 5). Stars with
  $V>14.5$ are plotted as smaller dots to keep the isochrones recognizable}
\label{cmdfield}
\end{figure*}

We derived colour-magnitude diagrams (CMDs) of each cluster by cutting out a
circular area with a radius of 100 pixels, corresponding to $34\arcsec$ or 8.3
pc, centered on the optical centre of each cluster. To derive the CMDs of
KMHK\,1019, which is the smallest of the three clusters, we adopted a smaller
radius of 80 pixels, corresponding to $27\farcs2$ or 6.6 pc.
An estimation by eye based on the star density plot (Fig. \ref{sl538dens}) 
suggests that no or almost no cluster stars are outside this area.  

All CMDs are plotted in Figs. \ref{cmdsl538} to \ref{cmdfield}. Each cluster
CMD has a wide blue main sequence and contains very few supergiants. The width
of the main sequence is caused in part by photometric errors, crowding (seeing
$1\farcs3$) and the presence of Be stars (see Sect. 5). The scarcity of
supergiants is well within the expected fluctuations for compact clusters with
few stars. 

Overplotted on the CMDs are the best fitting isochrones. We fitted the
isochrones such that the supergiants rather match the blue loops
than the quickly traversed subgiant branch. A distance modulus of 18.5 mag
(Westerlund 1997) was adopted. The metallicity of the young field population
of the LMC was found to be $[Fe/H]\simeq-0.3\pm0.2$~dex by various authors
(Russell \& Bessell 1989, Luck \& Lambert 1992, Russell \& Dopita 1992,
Th\'{e}venin \& Jasniewicz 1992). We therefore adopted Geneva isochrones with
Z=0.008 which corresponds to $[Fe/H]\simeq-0.3$~dex.    

Galactic field stars contaminate our observed area. Ratnatunga \& Bahcall
(1985) estimate the number of foreground stars towards the LMC, and in
Table \ref{foreground} we present their counts scaled to our total field of
view ($5\farcm8 \times 5\farcm8$).

\begin{table}
\caption[]{\label{foreground}Number of foreground stars towards the LMC
  calculated from the data of Ratnatunga \& Bahcall (1985), scaled to our
  field of view of $5\farcm8 \times 5\farcm8$}
\begin{tabular}{lccccc}
\hline
                    & \multicolumn{5}{c}{apparent visual magnitude range}\\
colour range        &13-15&15-17&17-19&19-21 &21-23 \\
\hline
$(B-V) < 0.8$       & 1.3 & 2.9 & 3.1 & 6.4  & 6.0\\
$0.8 < (B-V) < 1.3$ & 0.5 & 2.6 & 5.4 & 4.3  & 7.4\\
$1.3 < (B-V)$       & 0.1 & 0.6 & 3.7 & 13.1 & 29.3\\
\hline
\end{tabular}
\end{table}

{\it SL\,538}: We see two red and one blue supergiant in the $V$, $B-V$
CMD of SL\,538. 
The 16 Myr (solid line) isochrone fits all three supergiants well but also the
20 Myr (dotted line) isochrone fits the red supergiants very well.
We adopt an age of $18\pm2$ Myr. The same age is
found from isochrone fits to the $V$, $V-I$ CMD (Fig. \ref{cmdsl538}). All
isochrones are based on a reddening of $E_{B-V}=0.05$~mag.  

{\it NGC\,2006}: Four supergiants are located in this cluster, covering a
colour range from $B-V=0$ to 1.5 mag. 
Both CMDs (Fig. \ref{cmdn2006}) are fit quite well by isochrones with ages of
20 Myr (solid line) and 25 Myr (dotted line). We adopt an age of 22.5 Myr
($\pm2.5$ Myr) and a reddening of  $E_{B-V}=0.05$~mag.

{\it KMHK\,1019}: This cluster by far is the smallest one with the lowest
number of stars. 
Few data points are located in the red
clump and the red giant branch (RGB) region, and it is very likely that these
stars belong to an intermediate-age field star population, while the
supergiants, on which our age determination mainly relies, are located in the
cluster centre and thus we assume that they belong to the star cluster. The 
main sequence is sparse, especially in the upper part
brighter than $V\simeq16$ mag. The best fitting isochrones in both
CMDs result in an age of 16 Myr (Fig. \ref{cmdkmhk}). The reddening of
$E_{B-V}=0.05$~mag is the same for all fits.  

{\it The surrounding field}: The field population comprises a mixture of
ages. Apart from a blue main sequence and blue and red supergiants, which
represent the young field populations, the intermediate-age field population
of the LMC shows up through red giants and the pronounced red horizontal
branch clump. We are not able to distinguish between distinct young
populations, but the plotted isochrones represent ages which are supported by
corresponding supergiants.

The brightest blue supergiants and some of the brightest red supergiants are
represented by the 16 Myr isochrone (solid line). Also the 25 Myr (dotted) and
32 Myr (dashed) isochrones are supported by bright blue, yellow and red
supergiants. 

Several supergiants are traced by the 80 Myr isochrone. Note that the redder
main sequence stars at $V\simeq16$~mag are candidate Be stars (see
Sect. 5). One could easily mistake them for more evolved 
stars marking an additional field population with an age of 
approximately 100 Myr. The stellar density seems to be lower between 80 Myr
and 200 Myr which indicates a possible decrease in the field star formation
rate.  

Along the 200 Myr isochrone and below, the star density is increased along the
subgiant branch, again indicating enhanced star formation. 

$E_{B-V}=0.02$ mag is a lower limit to the reddening of the field star
populations and corresponds to the blue envelope of the main sequence(s). 

The youngest field population is part of LH\,77. Our derived age of
approximately 16 Myr is in good agreement with the findings of Braun et
al. (1997). 

\subsection{Comparison to earlier photometry}

Our study is the first age determination of SL\,538 and NGC\,2006 that is
based on CMDs. Previous studies derived ages based on surface
photometry, using different aperture sizes. Age determinations based on
integrated colours are less precise than age determinations based on CMDs:
Geisler et al. (1997) investigated the influence of a few bright stars on
the integrated light of intermediate age star clusters. They conclude that
fluctuations in the number of bright main sequence stars and red giants lead
to shifts in the integrated colours, which affect the age determination. 
In Table \ref{comphot} we present
a comparison to the integrated photometry of Bica et al. (1996), Bhatia
(1992) and our results.

Though several authors using surface photometry state that integrated colours
and thus the derived ages are largely independent from the aperture radius, 
Bica et al. (1996) and Bhatia (1992) derived quite different ages for SL\,538
and NGC\,2006:
Bica et al. (1996) used an aperture size of $50\arcsec$ and found NGC\,2006 to
be the older component of the cluster pair. This is in agreement with our
results. In contrast, Bhatia (1992) used an aperture radius of $33\arcsec$
and found SL\,538 to be slightly older than NGC\,2006. 
Our CMDs exclude ages as young as suggested by Bhatia's (1992) aperture
photometry. Bica et al.'s (1996) young age of only 0-10 Myr for SL 538 again
is clearly excluded by our data, while the wide age range of 10-30 Myr for
NGC 2006 includes our result of $22.5\pm2.5$ Myr.
Note that Bica et al.'s (1996) apertures are so large that stars belonging to
the neighbouring cluster are also included in their measurement.

\begin{table}
\caption[]{\label{comphot}Comparison of earlier age determination based on
  integrated colours and ours based on CMDs}
\begin{tabular}{lccc}
\hline
reference & aperture size & cluster & age\\
\hline\hline
                                    & & SL\,538 & 0--10 Myr\\
\raisebox{1.5ex}[-1.5ex]{Bica et al. (1996)} &
\raisebox{1.5ex}[-1.5ex]{$50\arcsec$} & NGC\,2006 & 10--30 Myr\\\hline
                                    & & SL\,538 & 12.6 Myr\\ 
\raisebox{1.5ex}[-1.5ex]{Bhatia (1992)} &
\raisebox{1.5ex}[-1.5ex]{$33\arcsec$} & NGC\,2006 & 7.9 Myr\\\hline
                                    & & SL\,538 & 18 Myr\\ 
\raisebox{1.5ex}[-1.5ex]{this work} &
\raisebox{1.5ex}[-1.5ex]{$34\arcsec$, CMD} & NGC\,2006 & 22.5 Myr\\ 
\hline
\end{tabular}
\end{table}

\section{Be stars in the clusters and the surrounding field}

\begin{figure}
\centerline{
\psfig{figure=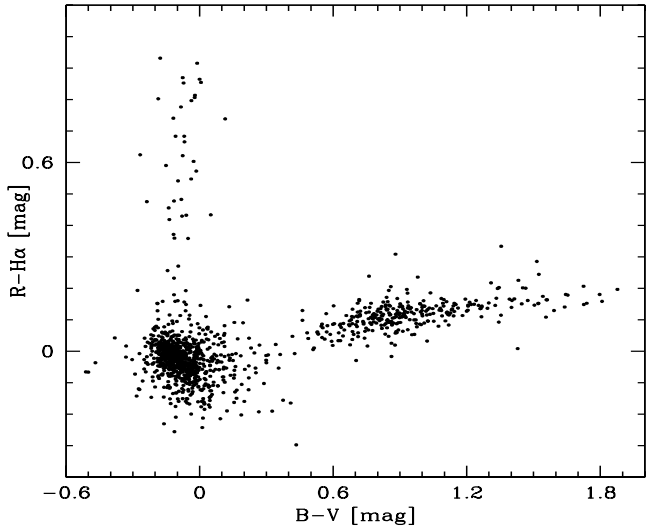,width=8.7cm,height=7.1cm}}
\vspace{-0.5cm}
\caption[]{Two-colour diagram to detect Be star candidates. $R$ serves as
  a continuum filter to detect stars bright in $H\alpha$ ($R-H\alpha>0.2$), 
  while $B-V$ serves as a temperature index to separate the blue stars from
  red giants and supergiants. For more information see text}
\label{beplot}
\end{figure}

Be stars are non-supergiant B stars with variable Balmer emission and infrared
excess originating in circumstellar disks.
Rapid rotation and red/infrared excess of Be stars lead to redder colours and
widened main sequences (see Grebel et al. 1996). The observed brightness and
colour of a rotating (Be) star depend on its rotational velocity and
inclination to the line of sight (rotational displacement fan, Collins \&
Smith 1985).  

Using the $R-H\alpha$ index to detect stars bright in $H\alpha$ (R serves
as continuum filter) and $B-V$ as a temperature index we can identify Be 
star candidates (Fig. \ref{beplot}). This method was first described in
Grebel et al. (1992, 1993). 
In Fig. \ref{beplot} a pronounced clump of data points at
$R-H\alpha\simeq0$ mag and $B-V\simeq0$ mag can be seen. These data points
represent blue main sequence stars and blue supergiants without
$H\alpha$ emission. To the red side of this clump, scattered around
$R-H\alpha\simeq0.1$ mag and extended over the whole $B-V$ colour range, stars
belonging to the RGB and red supergiants are visible in the two-colour
diagram. Red giants and red supergiants can have $H\alpha$ emission, and indeed
some data points are located at higher $R-H\alpha$ values
(up to $\simeq0.35$ mag). Our selection criteria for the candidate Be stars are
$B-V<0.2$ mag and $R-H\alpha>0.2$ mag. 

The widening of the main sequence of the CMDs due to Be stars can be clearly
seen in Figs. \ref{cmdsl538} to \ref{cmdfield}, where we marked the Be
star candidates with crosses. The effect is most pronounced in $V-I$ since 
the $I$ filter has the highest sensitivity to the infrared excess of Be stars.

In Fig. \ref{sl538} we marked these stars with circles. 
Our Be star candidates are concentrated at the location of the components of
the cluster pair whereas at the location of KMHK\,1019 only two such stars
are present. 

The impression that the Be star candidates are dominantly present in SL\,538
and NGC\,2006 is confirmed when considering the ratio of Be stars to B
stars. Since we do not have spectral classifications we simply considered 
the ratio of all B to Be stars within a magnitude interval of $V=14.2$ to 19.1
mag. These magnitudes correspond to the mean visual magnitudes, at LMC
distance, of B\,0\,III to B\,9\,V main sequence stars (Zorec \& Briot 1991,
Table 3).   
We find the following values: 

\begin{table}
\begin{tabular}{ll}
SL\,538:    & $N(Be)/N(B) = 0.123\,{}^{+0.134}_{-0.074}$\\
NGC\,2006:  & $N(Be)/N(B) = 0.120\,{}^{+0.145}_{-0.078}$\\
KMHK\,1019: & $N(Be)/N(B) = 0.053\,{}^{+0.233}_{-0.052}$\\
field:      & $N(Be)/N(B) = 0.019\,{}^{+0.022}_{-0.013}$
\end{tabular}
\end{table}

The errors are corresponding to $3 \sigma$ Gaussian errors and are
calculated using the confidence limits for small number statistics from
Gehrels (1986). The components of the double cluster show the same fraction
of Be stars, and the sixfold amount found in the surrounding field. The
difference between the ratios $N(Be)/N(B)$ of SL\,538 and the field is $0.104$
which is more than a $3 \sigma$-effect. The difference between SL\,538 and
KMHK\,1019 is $0.070$ which is less than a $2 \sigma$-effect according to the
upper confidence limit for KMHK\,1019. Thus, the Be star content of KMHK\,1019
may be comparable to the cluster pair.

\section{Summary and Conclusion}

Investigating the stellar density around the clusters (see
Fig. \ref{sl538dens}, right) we see no signs of increased star density
reaching from KMHK\,1019 towards the cluster pair, but we see  an enhanced
star density between SL\,538 and NGC\,2006. This may indicate a stellar bridge
and thus gravitational interaction between SL\,538 and NGC\,2006. The stars in
the bridge are main sequence stars and thus they may either belong to the
cluster {``}system{''} or to the young field star populations. However,
artificial star experiments showed that a stellar bridge may also be explained
(24\% probability for a {``}small{''} but only 6\% probability for a more
pronounced bridge) by statistical fluctuations in the field star density.

Fitting Geneva isochrones (Schaerer et al. 1993) to the CMDs we find the
following ages: SL\,538: $18\pm2$ Myr, NGC\,2006: $22.5\pm2.5$ Myr,
KMHK\,1019: 16 Myr, youngest field: 16 Myr.
The three clusters might have formed sequentially as part of the same GMC that
formed LH\,77.

Be stars are concentrated in SL\,538 and NGC\,2006, and both clusters show the
same ratio of $N(Be)/N(B)$. This is in agreement with Kontizas et al. (1993)
who investigated integrated IUE spectra and the distribution of spectral types
of stars and found that both clusters have similar stellar content.
Since Be stars are usually rapid rotators this may
indicate intrinsically higher rotational velocities in the components of the
cluster pair. The amount of Be stars detected in the surrounding field is
considerably lower.
 
An investigation of the IMF of the binary cluster candidate showed that
the IMF slopes agree with each other within the errors and are
compatible with the Salpeter value ($\Gamma = -1.35$, Salpeter 1955). Our
results for the IMF slopes are 
$\Gamma = -1.22 \pm 0.31$ for SL\,538 and $\Gamma = -1.27 \pm 0.32$ for
NGC\,2006. With these values we estimate the upper limits for the total
cluster masses to be $2300 \pm 1100 M_{\odot}$ (SL\,538) and $2300 \pm 1200
M_{\odot}$ (NGC\,2006). The similarity of the cluster masses is in agreement
with the findings from Kontizas et al. (1993). Let us assume that the cluster
pair indeed is a binary system: Following Kepler's third law and assuming
$4600 M_{\odot}$ for the total mass of the binary system and 13.3 pc for the
distance between the clusters we get an orbit period of  $\approx 47$
Myr. This would mean that the clusters so far have moved less than half of
an orbit since their formation.    

The similarity of properties (ages, Be star content, slope of the IMF and
masses) indicates possible joint formation and suggests small spatial
separation. Our results suggest that SL\,538 and NGC\,2006 are a 
true binary cluster, but a final proof is still missing. Radial velocity
measurements would help to understand the kinematics of the clusters and may
prove or disprove its possible binary nature, but such data are not yet
available. If the stellar bridge is real this may imply that the merger
process has begun. However, our artificial star experiments showed that we
cannot state whether this bridge is real or not.    

\acknowledgements

We would like to thank Prof. H. Els\"{a}sser for allocating time
at the MPIA 2.2m-telescope at La Silla during which our data were obtained,
Antonella Vallenari for introducing AD to the methods of completeness
correction, and Klaas S. de\,Boer and J\"{o}rg Sanner for a critical reading
of the manuscript. 
This work was supported by a graduate fellowship of the German Research
Foundation (Deutsche Forschungsgemeinschaft -- DFG) for AD through
the Graduiertenkolleg `The Magellanic System and Other Dwarf Galaxies' (GRK
118/2-96). EKG acknowledges support by  H.W. Yorke through grant 05 OR
9103 0 of the German Space Agency (DARA), travel support through GRK
118/2-96, and support by Dennis Zaritsky through NASA LTSA grant NAG-5-3501. 
    
This research has made use of NASA's Astrophysics Data System Abstract
Service and of the SIMBAD database operated at CDS, Strasbourg, France.

\end{document}